\documentstyle[preprint, aps]{revtex}

 \font\tenmsb=msbm10 scaled\magstep 1
   \font\sevenmsb=msbm7 scaled \magstep 1
   \font\faivemsb=msbm5 scaled \magstep 1
\newfam\msbfam
      \textfont\msbfam=\tenmsb
      \scriptfont\msbfam=\sevenmsb
      \scriptscriptfont\msbfam=\faivemsb
\def\Bbb#1{{\fam\msbfam #1}}

\newcommand{\be}{\begin{equation}}
\newcommand{\ee}{\end{equation}}
\newcommand{\Dlt}{\Delta}

\newcommand{\om}{\omega}
\newcommand{\ep}{\varepsilon}

\newcommand{\br}{{\bf r}}

\newcommand{\bt}{\beta}
\newcommand{\al}{\alpha}
\newcommand{\gm}{\gamma}
\newcommand{\Gm}{\Gamma}
\newcommand{\ra}{\rightarrow}

\newcommand{\lbd}{\lambda}
\newcommand{\prt}{\partial}

\tightenlines

\begin{document}
\draft
\title{Modified semiclassical approximation for trapped Bose gases}
\author{V.I. Yukalov}

\address{Institut f\"ur Theoretische Physik, \\
Freie Universit\"at Berlin, Arnimallee 14, D-14195 Berlin, Germany \\
and \\
Bogolubov Laboratory of Theoretical Physics, \\
Joint Institute for Nuclear Research, Dubna 141980, Russia}

\maketitle

\begin{abstract}

A generalization of the semiclassical approximation is suggested allowing
for an essential extension of its region of applicability. In particular,
it becomes possible to describe Bose-Einstein condensation of a trapped
gas in low-dimensional traps and in traps of low confining dimensions, for
which the standard semiclassical approximation is not applicable. The results
of the modified approach are shown to coincide with purely quantum-mechanical
calculations for harmonic traps, including the one-dimensional harmonic trap.
The advantage of the semiclassical approximation is in its simplicity and
generality. Power-law potentials of arbitrary powers are considered. Effective
thermodynamic limit is defined for any confining dimension. The behaviour
of the specific heat, isothermal compressibility, and density fluctuations
is analyzed, with an emphasis on low confining dimensions, where the usual
semiclassical method fails. The peculiarities of the thermodynamic
characteristics in the effective thermodynamic limit are discussed.

\end{abstract}


\section{Introduction}

Physics of Bose gases, exhibiting Bose-Einstein condensation, is
currently a subject of intensive investigations, both experimental
and theoretical (see, e.g., the review works [1--5]). A very efficient
method for describing Bose-Einstein condensation of trapped atoms is the
semiclassical approximation, which has been employed for considering the
properties of Bose-condensed ideal gases trapped in power-law potentials
[2,6--11]. The advantage of the semiclassical method is its simplicity,
as compared to quantum-mechanical calculations, and its generality allowing
for the treatment of power-law potentials of arbitrary powers, except those
cases, when the effective confining dimension becomes low. The latter happens
for the gas of low spatial dimensionality and when the power of the confining
potential is large. For instance, the semiclassical approximation predicts
the absence of Bose-Einstein condensation at finite temperature in a
one-dimensional harmonic trap [2,7,11], while this exists in a
quantum-mechanical picture [12].

In the present paper, a modification of the semiclassical approximation is
advanced, which makes it possible to essentially extend the region of its
applicability. Thus, the low-dimensional gases can be successfully described,
with confining potentials of arbitrary powers. The definition of an effective
thermodynamic limit for trapped atoms, known for the case of a
three-dimensional harmonic trap [3], can be generalized to arbitrary space
dimensionality and any confining powers. The specific-heat discontinuity at
the condensation point can now be studied for all those cases, for which the
standard approach does not work. The behaviour of isothermal compressibility
and the peculiarity of density fluctuations of a finite number of trapped
atoms, not studied earlier, are investigated.

Throughout the paper, the system of units is employed, where the Planck
constant $\hbar\equiv 1$ and the Boltzmann constant $k_B\equiv 1$ are set
to unity.

\section{Semiclassical Density of States}

Let us start with very briefly recalling the basic notions and introducing
the necessary notation that will be used in the following sections. We shall
consider the ideal Bose gas confined by means of a trapping potential
$U(\br)$. The Cartesian vector $\br=\{ r_\al\}$, with $\al=1,2,\ldots,d$,
is defined in a $d$-dimensional space, so that $d=1,2,\ldots$. The trapping
potential is assumed to be slowly varying in space, such that its
characteristic length $l_0$ be much larger than the thermal wavelength
$\lbd_T$,
\be
\label{1}
\frac{\lbd_T}{l_0} \ll 1 \; , \qquad \lbd_T \equiv
\sqrt{\frac{2\pi}{mT}}\; ,
\ee
where $m$ is atomic mass and $T$, temperature. Condition (1) can be rewritten
as an inequality for the characteristic frequency of the trapping potential
$\om_0$,
\be
\label{2}
\frac{\om_0}{T} \ll 1 \; , \qquad \om_0 \equiv \frac{1}{m l_0^2} \; .
\ee

Under condition (1) or (2), the semiclassical approach is applicable
[2,6--10], described by the density of states
\be
\label{3}
\rho(\ep) = \frac{(2m)^{d/2}}{(4\pi)^{d/2}\Gm(d/2)} \;
\int_{\Bbb{V}_\ep} \left [ \ep - U(\br)\right ]^{d/2-1}\; d\br \; ,
\ee
where $d\geq 1$ and the integration is over the available volume
$$
\Bbb{V}_\ep \equiv \{ \br|\; U(\br) \leq \ep \} \; .
$$
In particular, for a one-dimensional system, one has
\be
\label{4}
\rho(\ep) = \frac{\sqrt{2m}}{2\pi} \;
\int_{-x_\ep}^{x_\ep} \; \frac{dx}{\sqrt{\ep-U(x)}} \; ,
\ee
with the turning points $\pm x_\ep$ given by the equality $U(x_\ep)=\ep$.
Note that Eq. (4) follows exactly from Eq. (3).

The general form of a power-law confining potential can be represented as
\be
\label{5}
U(\br) = \sum_{\al=1}^d \; \frac{\om_\al}{2}\left |
\frac{r_\al}{l_\al}\right |^{n_\al} \qquad \left ( l_\al \equiv
\frac{1}{\sqrt{m\om_\al}}\right ) \; ,
\ee
where $\om_\al,l_\al$, and $n_\al$ are positive parameters. The
characteristic trapping frequency and potential length are given by the
geometric averages
\be
\label{6}
\om_0 \equiv \left ( \prod_{\al=1}^d \om_\al\right )^{1/d} \; , \qquad
l_0 \equiv \left ( \prod_{\al=1}^d l_\al \right )^{1/d} \; .
\ee

An important notion, arising for the confining potential (5), is the {\it
confining dimension}
\be
\label{7}
s \equiv \frac{d}{2} + \sum_{\al=1}^d \frac{1}{n_\al} \; .
\ee
The density of states (3), for potential (5), becomes
\be
\label{8}
\rho(\ep) = \frac{\ep^{s-1}}{\gm_d\Gm(s)} \; ,
\ee
where $\Gm(s)$ is a gamma-function. Form (8) is valid for any $d\geq 1$ and
$s\geq 1/2$. The parameter $\gm_d$ is
\be
\label{9}
\gm_d \equiv \frac{\pi^{d/2}}{2^s} \; \prod_{\al=1}^d
\frac{\om_\al^{1/2+1/n_\al}}{\Gm(1+1/n_\al)}
\ee
for $d\geq 1$. For instance, for $d=1$, it is
\be
\label{10}
\gm_1 = \frac{\sqrt{\pi}}{\Gm(1+1/n)} \left ( \frac{\om_0}{2}\right )^s
\qquad (d=1) \; .
\ee

In the case of harmonic traps, when $n_\al=2$, the confining dimension (7)
coincides with the spatial dimension $d=s$. Then $\gm_d=\om_0^d$ for all
$d\geq 1$, and the density of states is
$$
\rho(\ep) = \frac{\ep^{d-1}}{\Gm(d)\om_0^d} \qquad (d \geq 1) \; .
$$

\section{Modified Semiclassical Approximation}

Observable quantities, calculated with the density of states (8), are
expressed through the Bose-Einstein function $g_s(z)$, in which $s$ is
the confining dimension (7) and $z\equiv e^{\bt\mu}$ is fugacity, with
$\bt\equiv 1/T$ being inverse temperature and $\mu$, chemical potential.
Below the Bose condensation temperature $T_c$, one has $\mu\ra 0$ and
$z\ra 1$. The Bose-Einstein function $g_s(z)$ diverges as $z\ra 1$,
if $s\leq 1$. Those observable quantities that contain $g_s(1)$, with
$s\leq 1$, cannot be defined, because of the divergence of $g_s(1)$.
This imposes the limits of applicability for the semiclassical
approximation.

However the divergence of $g_s(1)$, for $s\leq 1$, is related to the usage
of the thermodynamic limit assuming an infinite system with an infinite
number of particles. For this case, integrations over momenta start from
zero, which implies that the minimal momentum is zero. But if the system
is finite, containing a finite number of particles $N$, though may be very
large, and in addition is confined in space by a trapping potential with
a characteristic length $l_0$, then the minimal momentum of a particle is
not zero, but rather is a finite quantity $k_{min}=1/l_0$. Respectively,
the minimal energy is $k_{min}^2/2m$, and the related dimensionless minimal
energy is
\be
\label{11}
u_{min} = \frac{\bt k_{min}^2}{2m} = \frac{\bt}{2ml_0^2}
= \frac{\om_0}{2T}\; .
\ee
The Bose-Einstein functions $g_n(z)$, arising in the process of calculating
physical quantities, are defined through integrals over the dimensionless
energy variable $u=\bt\ep$. For a finite confined system, the integration
should start from the minimal value (11), which yields
\be
\label{12}
g_n(z) = \frac{1}{\Gm(n)} \; \int_{u_{min}}^\infty \;
\frac{zu^{n-1}}{e^u-z}\; du \; .
\ee
The lower integration limit, according to inequality (2) is small,
\be
\label{13}
u_{min} =\frac{\om_0}{2T} \ll 1 \; ,
\ee
though it is not strictly zero. For $n>1$, when the integral (12) converges,
the value (13) is negligible and can be replaced by zero. However, for
$n\leq 1$, when the integral can diverge at $z\ra 1$, the lower integration
limit must be kept finite, being given by Eq. (11). In this way, we can
define the integral (12) for all $n$, which gives
$$
g_1(1) \cong \ln\; \frac{2T}{\om_0} \qquad (n=1) \; ,
$$
$$
g_n(1) \cong \frac{1}{(1-n)\Gm(n)} \left ( \frac{2T}{\om_0}\right )^{1-n}
\qquad (0 < n < 1) \; ,
$$
\be
\label{14}
g_0(1) \cong \frac{2T}{\om_0} \qquad (n=0) \; .
\ee
For the last equality in Eqs. (14), the relation
$$
\frac{\prt}{\prt z}\; g_n(z) = \frac{1}{z}\; g_{n-1}(z)
$$
is used, being valid for all $n$. For negative values $n<0$, the integral
(12) has the form of the second of Eqs. (14). Note that for all $n\geq 0$,
the values $g_n(1)$ are positive, while for $1/2\leq n<0$, these values
become negative, since then $\Gm(n)<0$.

The total number of particles can be represented as
\be
\label{15}
N=N_0 + \frac{T^s}{\gm_d} \; g_s(z) \; ,
\ee
where $N_0$ is the number of condensed atoms. From here, the critical
temperature $T_c$ follows as the temperature at which $\mu\ra 0$, $z\ra
1$, and $N_0\ra 0$, which results in
\be
\label{16}
T_c = \left [ \frac{\gm_d N}{g_s(1)} \right ] ^{1/s} \; .
\ee
For $s>1$, one has $g_s(1)=\zeta(s)$, with $\zeta(s)$ being a Riemann zeta
function. In the standard picture, $g_s(1)$ would diverge for all $s\leq 1$,
which would lead to the conclusion that then $T_c\ra 0$. That is, then a
finite condensation temperature would not exist for one-dimensional systems
trapped by a potential with the power $n\geq 2$, which includes the harmonic
potential [2,7,11]. But, as is explained above, for a finite confined system,
we have to take the value $g_n(1)$ given by Eqs. (14).

For $s=1$, which happens for a one-dimensional harmonic trap ($d=1$, $n=2$),
we find the critical temperature
\be
\label{17}
T_c = \frac{N\om_0}{\ln(2T_c/\om_0)} \qquad (s=1) \; ,
\ee
which is an immediate consequence of Eqs. (14) and (16). Equation (17)
exactly coincides with the equation for $T_c$ obtained in a purely
quantum-mechanical calculation [12]. Iterating Eq. (17), with taking account
of the inequality
$$
\frac{T_c}{\om_0} \ll e^{2N} \qquad (N\gg 1) \; ,
$$
we get the condensation temperature
\be
\label{18}
T_c = \frac{N\om_0}{\ln(2N)} \qquad (s=d=1,\; n=2)
\ee
for the one-dimensional harmonic trap.

In the case of one-dimensional traps, with confining potentials of powers
$n>2$, we can have the confining dimension in the interval $1/2\leq s<1$. In
such a case, the condensation temperature is
\be
\label{19}
T_c = \frac{\sqrt{\pi}(1-s)\Gm(s)}{2\Gm(1+1/n)}\; N\om_0 \qquad
(s < 1) \; .
\ee
Thus, we come to a conclusion that finite critical temperatures do exist
for one-dimensional traps for all powers of the confining potential.

Dealing with harmonic traps, one has $s=d$, so that for two-and
three-dimensional traps, the critical temperatures are
\be
\label{20}
T_c = \om_0 \left [ \frac{N}{\zeta(d)}\right ]^{1/d} \qquad
(s=d=2,3) \; .
\ee
And for a one-dimensional harmonic trap, $T_c$ is given by Eq. (18).

\section{Effective Thermodynamic Limit}

Effective thermodynamic limit for three-dimensional harmonic traps has
been defined earlier [3]. Here this notion is generalized for all spatial
dimensions $d\geq 1$ and arbitrary confining dimensions $s\geq 1/2$.

The basic idea in defining an effective thermodynamic limit is the common
agreement that, when the number of particles tends to infinity, $N\ra\infty$,
the extensive observable quantities must vary proportionally to $N$. One
of the main extensive observables is the internal energy $E$. Hence, the
thermodynamic limit in the most general sense can be understood as the
limit
\be
\label{21}
N\ra \infty\; , \qquad E\ra \infty\; , \qquad \frac{E}{N}\ra const\; ,
\ee
where a nonzero constant is assumed. For the internal energy, we have
\be
\label{22}
\frac{E}{N} = \frac{sg_{s+1}(z)}{N\gm_d}\; T^{s+1} \; .
\ee
Since $s\geq 1/2$, the value of $g_{s+1}(z)$ is always finite. This means
that the limit (21) can be rewritten as
\be
\label{23}
N\ra \infty\; , \qquad \gm_d\ra 0\; , \qquad N\gm_d\ra const \; .
\ee
This is the most general form of the thermodynamic limit, valid for all
power-law potentials in any spatial dimensionality.

Let us consider in more detail the case of unipower potentials, for which
$n_\al=n$. Then the confining dimension (7) becomes
\be
\label{24}
s =\left ( \frac{1}{2} + \frac{1}{n}\right ) d \; .
\ee
The coefficient (9) takes the form
\be
\label{25}
\gm_d = \frac{\pi^{d/2}}{\Gm^d(1+1/n)} \; \left (
\frac{\om_0}{2}\right )^s
\ee
where $\om_0$ is defined in Eq. (6). This shows that $\gm_d\propto \om_0^s$
for all $d\geq 1$. Thence, the limit (23) can be represented as the effective
thermodynamic limit
\be
\label{26}
N\ra\infty\; , \qquad \om_0\ra 0 \; , \qquad N\om_0^s\ra const \; .
\ee
The latter implies that
\be
\label{27}
\om_0 \propto N^{-1/s} \qquad (N\ra \infty) \; .
\ee
In a particular case of a three-dimensional harmonic trap, when $s=d=3$,
the limit (26) reduces to the known result [3]. But, generally, the limit
(26) is valid for all confining dimensions (24).

The condensation temperatures, found above, have sense for confined
systems with a finite, though very large, number of particles $N$. The
notion of the thermodynamic limit (26) allows one to analyze the behaviour
of $T_c$ as $N\ra\infty$. To this end, from Eq. (16) we have
\be
\label{28}
T_c \propto const \qquad (s>1) \; .
\ee
For the one-dimensional harmonic trap, Eq. (18) gives
\be
\label{29}
T_c \propto \frac{1}{\ln N} \ra 0 \qquad (s=1) \; .
\ee
And for very low confining dimensions $s$, from Eq. (19) we find
\be
\label{30}
T_c \propto N^{1-1/s} \ra 0 \qquad (s<1) \; .
\ee

Equations (28) to (30) demonstrate the behaviour of $T_c$ as $N\ra\infty$.
But for finite $N$, the corresponding values in Eqs. (16), (18), and (19)
can be finite and well defined.

\section{Specific Heat Discontinuity}

Specific heat, under a fixed number of particles, is given by the derivative
\be
\label{31}
C_N \equiv \frac{1}{N}\; \frac{\prt E}{\prt T} \; .
\ee
With the internal energy (22), for the temperatures above $T_c$, one has
\be
\label{32}
C_N = s(s+1) \; \frac{g_{s+1}(z)}{g_s(z)}\; - \; s^2\;
\frac{g_s(z)}{g_{s-1}(z)} \qquad (T > T_c) \; .
\ee
One may notice that for a three-dimensional uniform system, when $n\ra\infty$
and $s\ra 3/2$, Eq. (32) reduces to the known expression [13,14]. Below the
condensation temperature $T_c$, we get
\be
\label{33}
C_N = s(s+1)\; \frac{g_{s+1}(1)}{g_s(1)}\; \left (
\frac{T}{T_c}\right )^s \qquad (T < T_c) \; .
\ee
Expression (33) is finite and positive for all $s\geq 1/2$. Defining the
specific heat discontinuity at the critical point as
\be
\label{34}
\Dlt C_N \equiv C_N (T_c+0) - C_N(T_c-0) \; ,
\ee
we find
\be
\label{35}
\Dlt C_N = -s^2\; \frac{g_s(1)}{g_{s-1}(1)} \; .
\ee
In the standard approach, because of the divergence of $g_{s-1}(1)$ for
$s\leq 2$, the specific heat discontinuity (35) could be considered [11]
only for $s>2$. Here we extend the analysis for all $s$.

First, for higher confining dimensions, we have
\be
\label{36}
\Dlt C_N = -s^2\; \frac{\zeta(s)}{\zeta(s-1)} \qquad (s>2) \; ,
\ee
which agrees with the previous result [11]. For lower confining dimensions,
which could not be treated earlier, we obtain the following expressions:
\be
\label{37}
\Dlt C_N = -\; \frac{2\pi^2}{3\ln(2T_c/\om_0)} \; \qquad
(s=2) \; ,
\ee
\be
\label{38}
\Dlt C_N = -s^2 \zeta(s) (2 - s) \Gm(s-1)\left (
\frac{\om_0}{2T_c}\right )^{2-s} \qquad (1 < s < 2) \; ,
\ee
\be
\label{39}
\Dlt C_N = -\; \frac{\om_0}{2T_c} \; \ln\left (
\frac{2T_c}{\om_0}\right ) \qquad (s=1) \; ,
\ee
\be
\label{40}
\Dlt C_N = -\; \frac{s^2(2-s)\Gm(s-1)\om_0}{2(1-s)\Gm(s) T_c} \qquad
(s<1) \; .
\ee
For all $s\geq 1$, the specific heat jump is negative, which tells that
$$
C_N ( T_c -0) > C_N(T_c + 0) \qquad (s\geq 1) \; .
$$
But the sign of the discontinuity changes for $s<1$, demonstrating that
$$
C_N(T_c - 0) < C_N (T_c + 0) \qquad (s<1) \; .
$$
The jump (40) becomes positive, since $\Gm(s-1)$ is negative for
$1/2\leq s < 1$.

In the case of harmonic traps, the specific heat discontinuity
$\Dlt C_N$ is given by Eq. (39) for $d=1$, by Eq. (37) for $d=2$, and
for a  three-dimensional harmonic trap, one has
$$
\Dlt C_N = -\; \frac{54}{\pi^2}\; \zeta(3) \qquad (s=d=3) \; .
$$
For finite systems, the specific heat is always discontinuous at $T_c$.

\section{Density Fluctuations and Compressibility}

Density fluctuations in any statistical system can be quantified [13,15]
by the isothermal compressibility
\be
\label{41}
\kappa_T = \frac{\Dlt^2(\hat N)}{\rho TN} = \frac{1}{\rho^2}\left (
\frac{\prt\rho}{\prt\mu} \right ) \; ,
\ee
where $\rho$ is a mean particle density, and which is connected with the
number-of-particle dispersion
\be
\label{42}
\Dlt^2(\hat N) \equiv \; <\hat N^2 > \; - \; <\hat N>^2 \; .
\ee
For the latter, one has
\be
\label{43}
\Dlt^2(\hat N) =  T\; \frac{\prt N}{\prt\mu} =
\frac{TN}{\rho} \left ( \frac{\prt\rho}{\prt\mu}\right ) \; .
\ee
The total number of particles $N=N_0 + N_1$ is a sum of the number of
condensed atoms and the number of atoms outside the condensate. In a
gauge-symmetric grand canonical ensemble, the fluctuations of condensate
are known to be anomalous [16,17]. However, as is also perfectly known [16],
this is just an artifact that can be easily removed by breaking the gauge
symmetry. The most efficient way of gauge-symmetry breaking is by means of
the so-called Bogolubov shift [18,19], when the field operators of condensed
particles are replaced by the nonoperator quantities, representing the
condensate wave functions. In that way, the operator of the number of
condensed particles is replaced by its average, as a result of which only
the noncondensed particles contribute to the dispersion
\be
\label{44}
\Dlt^2(\hat N) = \Dlt^2(\hat N_1) = T\; \frac{\prt N_1}{\prt\mu} \; .
\ee
The necessity of introducing broken gauge symmetry in grand canonical
ensemble, in order to eliminate fictitious condensate fluctuations, was
discussed in great detail by ter Haar [16] and Hohenberg and Martin [20],
and carefully explained by Bogolubov [18,19]. The asymptotic exactness
of the Bogolubov shift was proven by Ginibre [21].

The number of noncondensed particles, according to Eq. (15), is
\be
\label{45}
N_1 = \frac{T^s}{\gm_d}\; g_s(z) \; .
\ee
From here, the dispersion (44) writes as
\be
\label{46}
\Dlt^2(\hat N_1) = \frac{T^s}{\gm_d}\; g_{s-1}(z) \; .
\ee
This, invoking relation (16), can be rewritten as
\be
\label{47}
\Dlt^2(\hat N_1) =  N\; \frac{g_{s-1}(z)}{g_s(1)} \left (
\frac{T}{T_c}\right )^s \; .
\ee
Involving the notion of the thermodynamic limit, particle fluctuations can
be classified into normal and anomalous. This is described more fully in
the review article [22]. When $\Dlt^2(\hat N_1)\propto N$, fluctuations
are called normal, while if $\Dlt^2(\hat N_1)\propto N^\al$, with $\al>1$,
they are termed anomalous.

At temperatures $T>T_c$, above the condensation point, one has $N_1=N$ and
$z<1$. For the confining dimension $s\geq 1$, particle fluctuations are
normal, since
\be
\label{48}
\Dlt^2(\hat N_1)\propto N \qquad (s\geq 1, T > T_c) \; .
\ee
Respectively, the compressibility (41) is finite for all $N\ra\infty$. For
lower confining dimensions $s<1$, we find
\be
\label{49}
\Dlt^2(\hat N_1) = \frac{zT}{(1-z)(1-s)\Gm(s-1)\gm_d}
\left ( \frac{2}{\om_0}\right )^{1-s} \; .
\ee
Then, because $\Gm(s-1)<0$ for $1/2<s<1$, the compressibility (41) becomes
divergent and negative,
\be
\label{50}
\kappa_T \propto - N^{-1+1/s} \qquad (s<1,\; T>T_c) \; .
\ee
Such a behaviour of the compressibility means that the system is
unstable.

Above the condensation temperature $T_c$, the trapped gas is stable only
for the confining dimensions $s\geq 1$. For harmonic traps, for which $s=d$,
the gas is stable in all spatial dimensions $d\geq 1$.

The situation is more interesting for the temperatures $T<T_c$ below the
condensation point. Then the dispersion (46) becomes
\be
\label{51}
\Dlt^2(\hat N_1) = \frac{T^s}{\gm_d}\; g_{s-1}(1) \; .
\ee
There exists a rich variety of different cases:
\be
\label{52}
\Dlt^2(\hat N_1) = N\; \frac{\zeta(s-1)}{\zeta(s)}
\left ( \frac{T}{T_c}\right )^s \qquad ( s > 2 ) \; ,
\ee
\be
\label{53}
\Dlt^2(\hat N_1) = \frac{N}{\zeta(2)}
\left ( \frac{T}{T_c}\right )^2\; \ln\left ( \frac{2T}{\om_0}\right )
 \qquad (s = 2) \; ,
\ee
\be
\label{54}
\Dlt^2(\hat N_1) = \frac{N}{(2-s)\zeta(s)\Gm(s-1)}
\left ( \frac{2T_c}{\om_0}\right )^{2-s}
\left ( \frac{T}{T_c}\right )^2 \qquad (1< s < 2) \; ,
\ee
\be
\label{55}
\Dlt^2(\hat N_1) = 2 \left ( \frac{T}{\om_0}\right )^2 \qquad
(s =1 ) \; .
\ee
And for $s<1$, the dispersion $\Dlt^2(\hat N_1)$ has the same form as in
Eq. (54). In the thermodynamic limit, we find
$$
\Dlt^2(\hat N_1) \propto N \qquad ( s > 2)\; ,
$$
$$
\Dlt^2(\hat N_1) \propto N\ln N \qquad ( s = 2)\; ,
$$
$$
\Dlt^2(\hat N_1) \propto N^{2/s} \qquad (1< s < 2)\; ,
$$
$$
\Dlt^2(\hat N_1) \propto N^2 \qquad ( s = 1 )\; ,
$$
\be
\label{56}
\Dlt^2(\hat N_1) \propto - N^{2/s} \qquad ( s < 1) \; .
\ee
Respectively, the behaviour of the compressibility is
$$
\kappa_T \propto const \qquad (s > 2) \; ,
$$
$$
\kappa_T \propto \ln N \qquad (s = 2) \; ,
$$
$$
\kappa_T \propto N^{2/s-1} \qquad (1< s < 2) \; ,
$$
$$
\kappa_T \propto N \qquad (s = 1) \; ,
$$
\be
\label{57}
\kappa_T \propto - N^{2/s-1} \qquad (s < 1) \; .
\ee

These equations demonstrate that the fluctuations are anomalous for all
$s\leq 2$. A negative compressibility for $s<1$ implies strong instability
of the system. For the confining dimensions in the interval $1\leq s\leq 2$,
the compressibility is positive but displays nonthermodynamic behaviour
diverging in the thermodynamic limit. An actual divergence of $\kappa_T$
happens only for infinite systems, when $N\ra\infty$, which would imply
instability. When one deals with finite system, with a large, though finite,
number of particles $N\gg 1$, then the compressibility does not really
diverge, but becomes very large. This means that there exist very strong
density fluctuations in the system. Such strong fluctuations arise in the
case of the low confining dimension $s\leq 2$, which corresponds to large
confining powers $n_\al$ and low spatial dimensions $d$. Fluctuations are
known to be strong in low-dimensional systems [23,24]. For higher confining
dimensions $s>2$, fluctuations are always normal, so that for
\be
\label{58}
\frac{d}{2} + \sum_{\al=1}^d \; \frac{1}{n_\al} > 2
\ee
the system is stable.

Since the standard traps are usually harmonic, let us pay a special attention
to the harmonic confining potentials, when $s=d$ and $\gm_d=\om_0^d$. Then
the dispersion (51) becomes
\be
\label{59}
\Dlt^2(\hat N_1) = g_{d-1}(1)\left (
\frac{T}{\om_0}\right )^d \; .
\ee
For different spatial dimensions, we find
$$
\Dlt^2(\hat N_1) = 2 \left (\frac{T}{\om_0}\right )^2 \qquad
(d = 1) \; ,
$$
$$
\Dlt^2(\hat N_1) = \left (\frac{T}{\om_0}\right )^2
\ln \left ( \frac{2T}{\om_0}\right ) \qquad (d = 2) \; ,
$$
\be
\label{60}
\Dlt^2(\hat N_1) = N \frac{\pi^2}{6\zeta(3)}  \left ( \frac{T}{T_c}
\right )^3 \qquad (d = 3) \; .
\ee
The last expression agrees with the corresponding result by Politzer
[25] for a three-dimensional harmonic trap. All other formulas of this
section are new.

Dispersions (60) in the thermodynamic limit behave as
$$
\Dlt^2(\hat N_1) \propto N^2 \qquad (d = 1) \; ,
$$
$$
\Dlt^2(\hat N_1) \propto N \ln N \qquad (d = 2) \; ,
$$
$$
\Dlt^2(\hat N_1) \propto N \qquad (d = 3) \; .
$$
So that the dispersions for $d=1$ and $d=2$ are anomalous. The related
compressibilities possess the limits
$$
\kappa_T \propto N \qquad (d = 1) \; ,
$$
$$
\kappa_T \propto \ln N \qquad (d = 2) \; ,
$$
$$
\kappa_T \propto const \qquad (d = 3) \; .
$$
Anomalous values of the compressibilities for low-dimensional harmonic
traps signify instability caused by the existence of very strong fluctuations
in such traps.

\section{Conclusion}

The semiclassical approximation is generalized, which makes it possible
to essentially extend the region of its applicability. Bose-Einstein
condensation in traps of low confining dimensions is described, for which
the standard approach could not be used. Trapping potentials of arbitrary
power laws are considered. Specific-heat discontinuities and isothermal
compressibilities are analyzed. Effective thermodynamic limit is defined
for any spatial dimension and for arbitrary powers of confining potentials.
It is shown that the modified semiclassical method for harmonic traps yields
the results coinciding with those obtained by means of quantum-mechanical
calculations, when these are available.

\vskip 5mm

{\bf Acknowledgement}

\vskip 2mm

I am grateful to the German Research Foundation for the Mercator
Professorship.

\end{document}